\documentclass[conference]{IEEEtran}
\usepackage{cite}
\usepackage{amsmath,amssymb,amsfonts}
\usepackage{algorithm}
\usepackage{algpseudocode}
\usepackage{booktabs}
\usepackage{graphicx}
\usepackage{multirow}
\usepackage{textcomp}
\usepackage{xcolor}
\usepackage[a4paper, total={184mm,239mm}]{geometry}
\def\BibTeX{{\rm B\kern-.05em{\sc i\kern-.025em b}\kern-.08em
    T\kern-.1667em\lower.7ex\hbox{E}\kern-.125emX}}

\DeclareMathOperator{\Root}{root}
\DeclareMathOperator{\Parent}{parent}
\DeclareMathOperator{\Growth}{growth}
\DeclareMathOperator{\Active}{active}
\DeclareMathOperator{\Token}{token}
\DeclareMathOperator{\Parity}{parity}
\DeclareMathOperator{\Corr}{corr}
\DeclareMathOperator{\Boundary}{boundary}
\DeclareMathOperator{\Defect}{defect}

\begin{document}

\title{Amatlas: A 2D Array Architecture for Union-Find Decoding Without Global Coordination}

\author{
\IEEEauthorblockN{
Takuya Kasamura,
Leo Itoh,
Seita Tsuboi,
Hidetsugu Irie,
and Junichiro Kadomoto
}
\IEEEauthorblockA{
\textit{Graduate School of Information Science and Technology, The University of Tokyo}, Tokyo, Japan\\
\{kasamura, itoh, tsuboi, irie, kadomoto\}@mtl.t.u-tokyo.ac.jp
}
}

\maketitle

\begin{abstract}
Real-time decoding of the surface code is essential for fault-tolerant quantum computing, and scaling the decoder to large code distances in hardware remains a challenge.
The locality of Union-Find (UF) decoding makes it a leading candidate for such a decoder in distributed hardware.
Prior distributed UF decoders, however, map the three-dimensional decoding graph onto an array of processing elements (PEs) under a central controller that synchronizes all PEs at every stage, and a three-dimensional array of PEs scales poorly on planar hardware.
This paper presents Amatlas, a two-dimensional PE array for UF decoding that operates without a central controller.
Amatlas assigns one PE to each stabilizer on the two-dimensional qubit lattice and stores the states of all measurement rounds in the local memory of the PE, so every link connects adjacent PEs in the array.
Every PE advances on every clock cycle without array-wide synchronization, and the clock is the only global signal of the array.
Under circuit-level noise, the RTL implementation of Amatlas fits a VCU129 FPGA up to a code distance of 25, where the average decoding time is 462\,ns per round at a physical error rate of 0.1\%.
Amatlas achieves a threshold of 0.72\%, compared with 0.78\% reported for UF decoding under the same noise model.
\end{abstract}


\section{Introduction}
\label{sec:intro}

Quantum computing has emerged as a promising approach to classically intractable problems.
However, the high susceptibility of quantum states to physical errors remains a major obstacle to its large-scale realization.
Fault-tolerant quantum computing (FTQC) therefore relies on quantum error correction (QEC), which encodes each logical qubit into multiple physical qubits and continuously detects and corrects errors.

The surface code~\cite{Dennis2002,Kitaev2003,Fowler2012} is a leading candidate for large-scale FTQC because of its high error threshold and nearest-neighbor connectivity.
Errors are detected by repeated stabilizer measurements, and the outcomes, called a syndrome, are processed by a classical decoder to infer corrections.
Since syndromes are continuously generated during computation, decoding must operate concurrently with syndrome extraction at approximately 1\,\textmu s per round~\cite{GoogleQuantumAI2025}.

For real-time decoding, the Union-Find (UF) algorithm~\cite{Delfosse2021} is attractive for its favorable accuracy and low complexity.
The algorithm organizes detected errors (defects) into clusters and repeatedly grows and merges them.
Since these operations involve only neighboring nodes, UF decoding is suitable for distributed hardware implementation.

However, this locality does not eliminate the need for global coordination in conventional UF decoding.
Each growth step completes in a fixed number of cycles, whereas merging requires cluster information to propagate repeatedly until convergence.
The next growth step must wait for confirmation of convergence from across the decoder, introducing global stage synchronization into an otherwise local algorithm.

Existing hardware implementations of UF decoding take different positions on this coordination.
In a centralized architecture, the Collision Clustering (CC) decoder~\cite{Barber2025} processes cluster operations sequentially in a single engine, achieving high resource efficiency.
However, the average decoding time increases with the number of defects and exceeds 1\,\textmu s per round beyond \(d=21\) at a physical error rate of 0.1\%.

In a distributed architecture, Helios~\cite{Liyanage2024} and a recent FPGA implementation~\cite{Tahaab2026} of Snowflake~\cite{Chan2026} assign one processing element (PE) to every node of the three-dimensional decoding graph under a central controller.
This node-level parallelism enables fast decoding, but the resulting three-dimensional array scales poorly on two-dimensional hardware.
Under circuit-level noise, Helios achieves 23.7\,ns per round but reaches only \(d=17\) on a large FPGA.
Although Helios-\(n\)~\cite{Liyanage2024} and the Local Clustering Decoder (LCD)~\cite{Ziad2025} fold the graph onto a two-dimensional array, a central controller must still confirm the completion of merging across all PEs before the next stage begins.
The centralized coordination identified above persists and constrains scalability.

This paper presents Amatlas, a natively two-dimensional distributed architecture for UF decoding without global coordination.
Amatlas assigns one PE to each stabilizer and stores the decoding state across measurement rounds in its local memory, so every inter-PE link connects adjacent PEs.

The contributions of this paper are as follows.

\begin{itemize}
\item A two-dimensional array architecture for UF decoding in which one PE per stabilizer stores the state of all measurement rounds in local memory.
\item The elimination of convergence detection and the central controller, based on an invariant that guarantees a valid correction regardless of the timing of growth, so that growth proceeds on a fixed schedule and the clock is the only global signal of the array.
\item An RTL implementation on the same FPGA as Helios, which fits up to \(d=25\) with average decoding time below 1\,\textmu s per round at every implemented distance and a threshold of approximately 0.7\% under circuit-level noise, compared with CC, Helios, Snowflake, and LCD.
\end{itemize}

\section{Background and Related Work}

\subsection{Surface-Code Decoding}

\begin{figure}[tbp]
  \centering
  \centerline{\includegraphics[width=\linewidth]{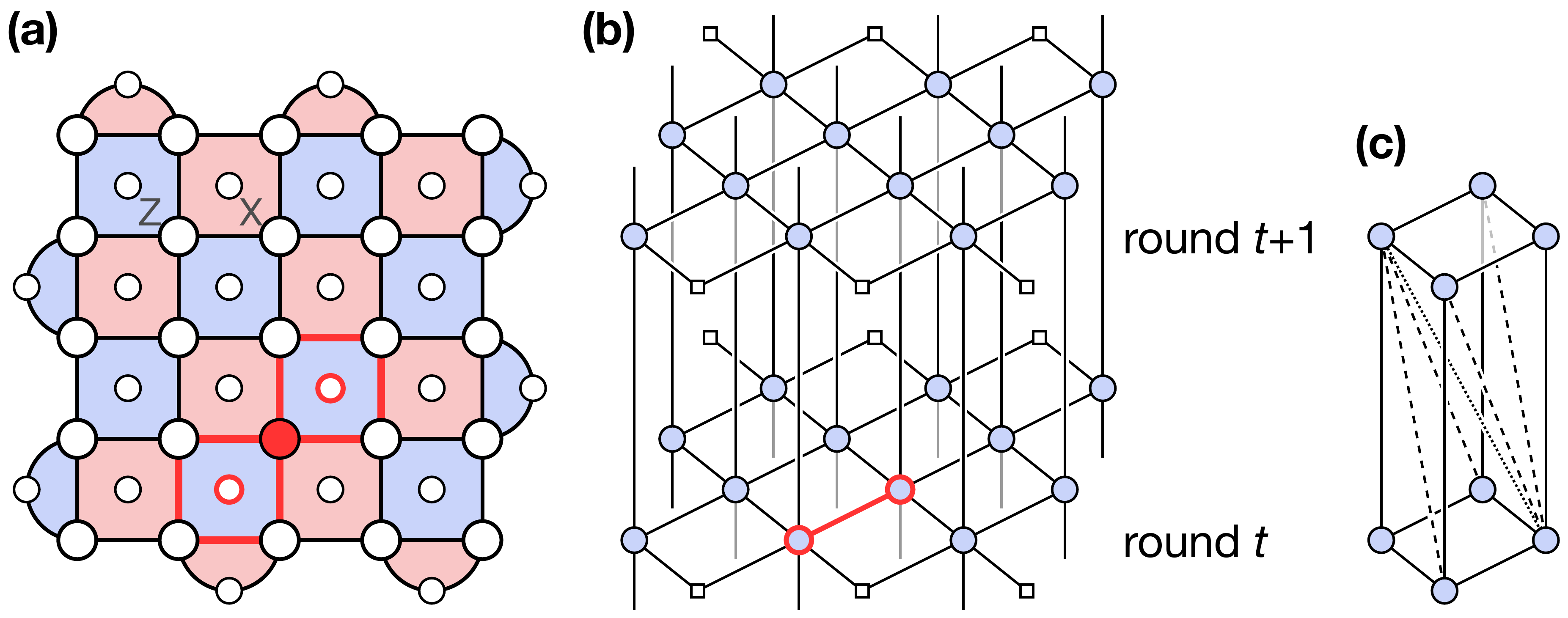}}
  \caption{Surface code and decoding graph. (a) Rotated surface code with \(d=5\). Blue and red plaquettes are Z-type and X-type stabilizers, and circles at their corners and centers are data and ancilla qubits, respectively. A bit flip on the red data qubit flips the two Z-type stabilizers outlined in red. (b) Decoding graph of the Z-type stabilizers under phenomenological noise. Squares are virtual boundary nodes. (c) Edges of one cell of the decoding graph over two rounds. Circuit-level noise adds the dashed diagonal edges and the dotted hook edge.}
  \label{fig:surface_code}
\end{figure}

The surface code consists of data qubits and ancilla qubits, as Fig.~\ref{fig:surface_code}(a) shows for \(d=5\), where the code distance \(d\) is the minimum number of physical errors that cause a logical error.
Logical information is encoded in the data qubits, and each ancilla qubit measures the parity of its neighboring data qubits.
This parity check is called a stabilizer.
The code detects bit flips and phase flips with Z-type and X-type stabilizers, respectively.
Since a measurement itself is subject to error, the stabilizers are measured repeatedly, and the outcomes of each round form the syndrome.
An error on a data qubit flips the adjacent stabilizers, whereas a measurement error flips one outcome in one round.
In either case, the error appears as a change of an outcome between consecutive rounds, called a defect.

The decoder infers the underlying errors from defects on separate decoding graphs for the two stabilizer types~\cite{Dennis2002}, and Fig.~\ref{fig:surface_code}(b) shows the graph of one type.
Each node corresponds to one stabilizer at one round, and each edge corresponds to a single error event.
Edges within a round represent data-qubit errors, and edges between rounds represent measurement errors.
The graph with only these two kinds of edges corresponds to phenomenological noise.
The more realistic circuit-level noise accounts for every gate of the measurement circuit and adds diagonal edges between stabilizers that differ in both position and round, as Fig.~\ref{fig:surface_code}(c) shows.
An error at the code boundary flips a single stabilizer and is represented by an edge to a boundary node.
The decoder selects chains of edges that connect each defect to another defect or the boundary, and the selected edges constitute the correction.
Decoding succeeds if the errors and the correction together preserve the logical information, and the probability of failure is the logical error rate.
The logical error rate decreases with \(d\) when the physical error rate \(p\) is below a threshold that depends on the decoder~\cite{Dennis2002}.

\subsection{Union-Find Decoding}

A decoder estimates the most likely set of errors that explains the observed defects.
Minimum-weight perfect matching (MWPM)~\cite{Edmonds1965,Higgott2025} finds this set exactly on the decoding graph with edges weighted by the error probabilities, but its computational cost increases rapidly with the number of defects.
Among the approximate decoders that reduce this cost~\cite{Ueno2021,Overwater2022}, the Union-Find (UF) decoder~\cite{Delfosse2021} has been widely studied, since it achieves almost-linear complexity with an error threshold close to that of MWPM.

In UF decoding, every defect initially forms a cluster of one node.
In each growth step, every cluster that contains an odd number of defects extends by half an edge in every direction.
Two clusters merge when their growth meets on an edge.
A cluster becomes inactive and stops growing once it contains an even number of defects or touches a boundary node, since its defects can then be paired within the cluster.
Once every cluster is inactive, the correction is obtained within each cluster by peeling a spanning tree~\cite{Delfosse2020}.
Because the uniform growth of clusters does not reflect the edge weights, UF decoding is less accurate than MWPM decoding.

\subsection{Hardware Union-Find Decoders}

UF decoders execute growth and merging as alternating stages and detect the end of merging by collecting a convergence flag from every node.
The Collision Clustering (CC) decoder~\cite{Barber2025} requires no such coordination, since it processes clusters sequentially in a single engine that holds all clusters in a global table.
However, the decoding time per round increases with the number of defects, which is proportional to \(pd^2\).

Helios~\cite{Liyanage2024} assigns one PE to every node of the three-dimensional graph, and a central controller collects a busy flag from every PE in every cycle to advance the stages.
It achieves an average decoding time of 11.5\,ns per round at \(d=21\) on a VCU129 FPGA under phenomenological noise, whereas circuit-level noise doubles the links per PE and limits the implementation to \(d=17\).
The Helios authors identify ``the architectural mismatch between the 3-D design of Helios and the 2-D structure of an FPGA'' as a fundamental limitation on scalability~\cite{Liyanage2024}.
A time-multiplexed variant (Helios-\(n\)) uses multi-context PEs~\cite{Heer2023} to fold \(n\) rounds onto each PE at a proportional cost in decoding time.
Its evaluation is limited to phenomenological noise, and no extension to circuit-level noise has been reported.

Macar and Actis~\cite{Chan2023} are distributed UF decoders with one node per PE, evaluated by emulation.
Macar computes cluster parities by moving defects toward the cluster root, and Actis relays the controller signals through neighboring nodes at a cost of 88\% of the runtime at \(d=25\).
Snowflake~\cite{Chan2026} additionally lets the moving defects flip the edges they traverse, so that the flipped edges constitute the correction without a peeling stage, yet its controller still waits until every node becomes idle before advancing the stage.
Its FPGA implementation~\cite{Tahaab2026} reaches \(d=9\) on a mid-range Kintex UltraScale+ FPGA.
The authors propose a two-dimensional array with one PE per stabilizer as future work, with a controller that loads slices of the graph from external memory, and provide pseudocode only.

The Local Clustering Decoder (LCD)~\cite{Ziad2025} also reduces the array to two dimensions.
Its array spans one spatial dimension and the temporal dimension of the graph.
LCD reaches \(d=17\) in under 1\,\textmu s per round on a VU19P FPGA under circuit-level noise with leakage, and its central controller still advances the stages only after every PE has reported that it is idle.

\section{Amatlas Architecture}

\subsection{Overview}

\begin{figure}[tbp]
  \centering
  \centerline{\includegraphics[width=\linewidth]{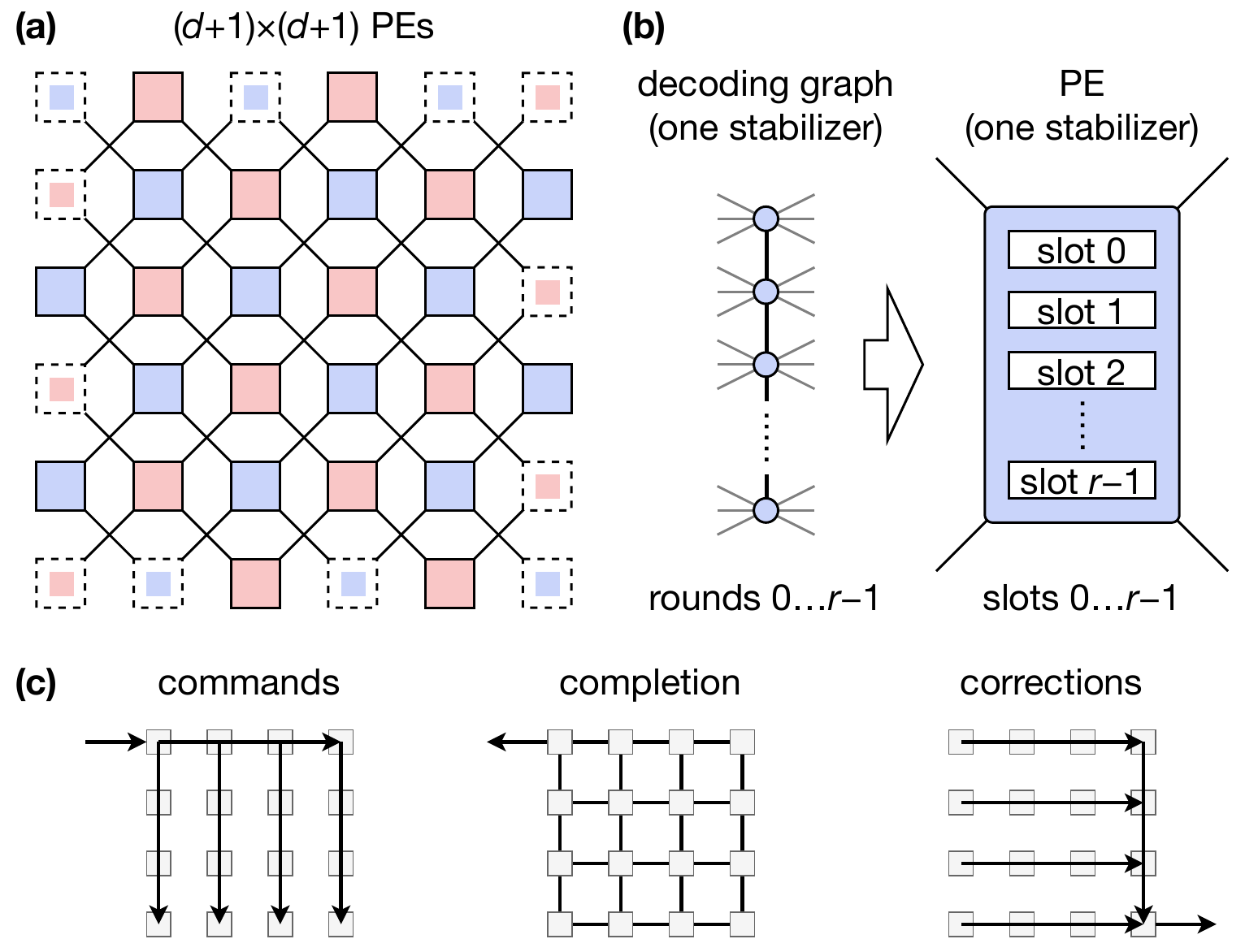}}
  \caption{Amatlas array for \(d=5\). (a) Identical PEs, one per node of the two-dimensional stabilizer lattice. The two colors distinguish the stabilizer types, and dashed squares indicate PEs without assigned stabilizers. Every link connects diagonally adjacent PEs. (b) The nodes of one stabilizer over \(r\) rounds occupy slots in the state memory of its PE. (c) Three chains connecting the array to the host. Commands configure the PEs and load the defects, completion reports the end of decoding, and corrections carry the output.
}
  \label{fig:array}
\end{figure}

Amatlas is an array of \((d+1)\times(d+1)\) PEs, one per node of the two-dimensional lattice of stabilizers.
Fig.~\ref{fig:array}(a) shows the array for \(d=5\).
X-type and Z-type stabilizers are arranged on this lattice in a checkerboard pattern, and a single array decodes both types.
The temporal dimension of the decoding graph is kept in memory rather than the PE array, as shown in Fig.~\ref{fig:array}(b).
Each PE holds the decoding state of its stabilizer in a local state memory of \(r\) slots, where \(r\) is the number of syndrome rounds in the decoding graph.
A node of the decoding graph is therefore identified by a PE and a slot.

Each PE has four links to its diagonally adjacent PEs, which host stabilizers of the same type.
A link carries three lanes for the edges to that neighbor in the same round and the two adjacent rounds.
All links end in a register at the receiving PE, and no combinational path crosses more than one PE.

Decoding proceeds in sweeps.
In one sweep every PE processes its \(r\) slots in sequence, and all PEs process the same slot at the same time.
Cluster growth is applied once every \(K\) sweeps, where the growth period \(K\) is a design parameter.

\subsection{Union-Find Decoding Without Global Coordination}

\label{subsec:uf}

Amatlas employs a token-based formulation of UF decoding that transports each defect as a token toward the root of its cluster and computes the parity of the cluster at the root.
This formulation follows Macar~\cite{Chan2023} and Snowflake~\cite{Chan2026}, although those decoders apply growth only after merging has converged.
Amatlas instead applies growth on a fixed schedule regardless of the progress of merging, and the invariant introduced below guarantees that the correction remains valid.
Each node holds the identifier of its cluster (\(\Root\)), the neighbor toward the root (\(\Parent\)), a growth level from 0 to 2 (\(\Growth\)), an activity flag (\(\Active\)), a token that carries a defect (\(\Token\)), the parity held as a root (\(\Parity\)), and one correction bit per incident edge (\(\Corr\)).

Algorithm~\ref{alg:node} lists the update of one node in one sweep in the three stages of the PE, namely gossip, select, and commit, followed by the growth step.
In the gossip stage, the node exchanges its root, growth level, and activity flag with the neighbors, and we call this information the gossip.
In this stage, the node also receives the tokens sent in the previous sweep.
The select stage determines the smallest root among the neighbors reached through fully grown edges, where the boundary is treated as a neighbor with the smallest identifier and zero growth.
The commit stage adopts the selected root and advances the token by one edge toward the parent, or absorbs the token at the root.
A token sent to the parent is processed at the next update of the parent, and the boundary absorbs every token sent to it.
The first growth step is applied to every active node at the end of the first sweep.
Growth is then repeated once every \(K\) sweeps, without waiting for merging to converge.

\begin{algorithm}[t]
\caption{Update of node \(v\) in one sweep}
\label{alg:node}
\small
\begin{algorithmic}[1]
\Statex \textsc{Gossip}
\State exchange \((\Root, \Growth, \Active)\) with all neighbors
\ForAll{tokens received from a neighbor \(u\)}
  \State \(\Token(v)\gets\Token(v)\oplus 1\); flip \(\Corr(v,u)\)
\EndFor
\If{\(\Root(v)=\Boundary\)} \(\Active(v)\gets 0\)
\ElsIf{\(\Root(v)=v\)} \(\Active(v)\gets \Parity(v)\)
\Else\ \(\Active(v)\gets \Active(\Parent(v))\)
\EndIf
\Statex \textsc{Select}
\State \(\mathcal{C} \gets \{(\Root(u), u) : \Growth(u)+\Growth(v)\ge 2\}\)
\State \((\Root^\ast, \Parent^\ast) \gets \min \mathcal{C}\)
\Statex \textsc{Commit}
\If{\(\mathcal{C}\neq\emptyset\) and \(\Root^\ast < \Root(v)\)}
  \State \((\Root(v), \Parent(v)) \gets (\Root^\ast, \Parent^\ast)\)
  \State \(\Token(v)\gets\Token(v)\oplus\Parity(v)\); \(\Parity(v)\gets 0\)
\EndIf
\If{\(\Token(v)=1\)}
  \State \(\Token(v)\gets 0\)
  \If{\(\Root(v)=v\)} \(\Parity(v)\gets\Parity(v)\oplus 1\)
  \Else\ flip \(\Corr(v,\Parent(v))\); send a token to \(\Parent(v)\)
  \EndIf
\EndIf
\Statex \textsc{Grow} (at the end of every \(K\)th sweep)
\If{\(\Active(v)=1\) and \(\Growth(v)<2\)} increment \(\Growth(v)\)
\EndIf
\end{algorithmic}
\end{algorithm}

The correctness of Amatlas rests on a residual-syndrome invariant.
A token that traverses an edge flips the correction bit of that edge at both ends.
The correction bits thus record the trajectories of the tokens, and no peeling stage is needed.
Let \(\Defect(v)\) denote the defect bit of node \(v\) at the start of decoding.
For every node \(v\) and at every clock cycle,
\begin{equation}
\Defect(v) \oplus \bigoplus_{u} \Corr(v,u) = \Token(v) \oplus \Parity(v),
\label{eq:invariant}
\end{equation}
where \(u\) ranges over the neighbors of \(v\).
The left-hand side is the syndrome that would remain if the current correction were applied, and the invariant states that this residual syndrome is located exactly at the tokens and the root parities.

The invariant holds initially, as all correction bits are zero and the token of every node equals its defect bit.
Every operation of Algorithm~\ref{alg:node} preserves it, since each operation flips either a token together with a root parity or a token together with the correction bit of the traversed edge, and growth changes none of the variables in \eqref{eq:invariant}.
Whenever no token and no root parity remain, the residual syndrome is zero at every node, and the correction bits form a valid correction.
The same relation is implicit in Definition~2 of Snowflake~\cite{Chan2026}, where the pushing of a defect is defined as the flipping of the traversed edge together with the corresponding update of the syndrome.
We state this property explicitly as an invariant.

With this invariant, growth requires no confirmation that merging has converged, and merging runs continuously alongside growth.
If a node grows before an ongoing merge has converged, the excess growth affects the accuracy but never the validity of the correction.
A small \(K\) grows clusters before the information of their merges has propagated and lowers the accuracy, whereas a large \(K\) increases the decoding time.
We set \(K=d+1\) and evaluate this choice in Section~\ref{subsec:accuracy}.

\subsection{Processing Element Microarchitecture}

\label{subsec:pe}

\begin{figure}[t]
  \centering
  \centerline{\includegraphics[width=\linewidth]{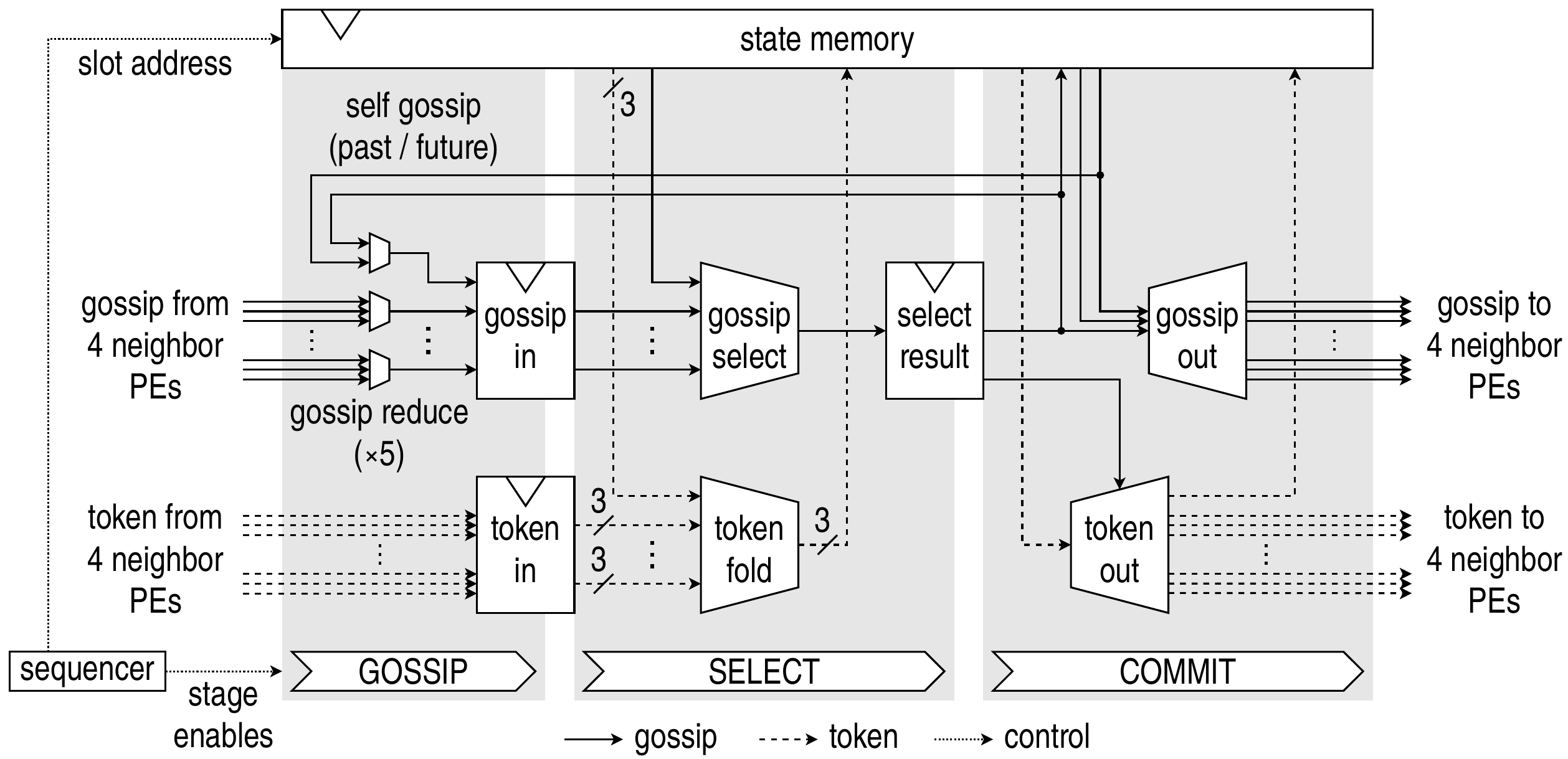}}
  \caption{Microarchitecture of one PE. The gossip, select, and commit stages are implemented as pipeline stages separated by registers, and each stage is traversed by a gossip path and a token path. The PE contains a state memory of \(r\) slots, four links to the diagonally adjacent PEs with three lanes each, and a sequencer that generates the slot address and the stage enables.}
  \label{fig:pe}
\end{figure}

Each PE executes Algorithm~\ref{alg:node} for its \(r\) slots in the three pipeline stages shown in Fig.~\ref{fig:pe}, which are separated by registers and each traversed by a gossip path and a token path.
The gossip path reduces the gossip from each link to one candidate, selects the smallest root among the candidates, and writes the result back to the links and the state memory.
The token path delivers the tokens arriving over the links to their destination slots and sends forwarded tokens to the adjacent PEs.

\begin{figure}[t]
  \centering
  \centerline{\includegraphics[width=\linewidth]{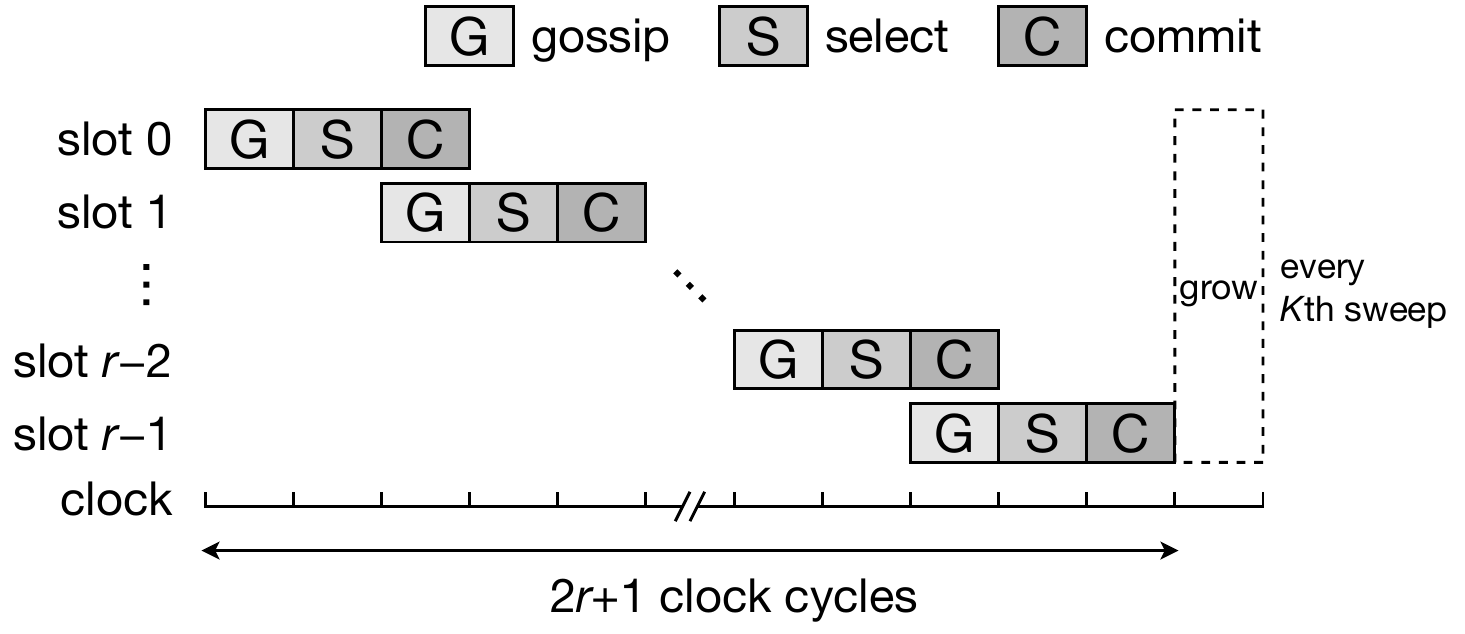}}
  \caption{Timing of one sweep. Each slot passes through the gossip (G), select (S), and commit (C) stages. The gossip stage of a slot overlaps with the commit stage of the previous slot, so a sweep of \(r\) slots takes \(2r+1\) cycles. Growth is applied to all slots in one additional cycle at the end of every \(K\)th sweep.}
  \label{fig:timing}
\end{figure}

The timing of one sweep is shown in Fig.~\ref{fig:timing}.
Each slot passes through the three stages in three clock cycles, and the gossip stage of a slot overlaps with the commit stage of the previous slot, so a sweep of \(r\) slots takes \(2r+1\) clock cycles.
Growth is applied to all slots in one additional cycle at the end of every \(K\)th sweep.
Within a sweep, the gossip propagates through many slots in the direction of the scan but only one slot against it.
Consecutive sweeps therefore visit the slots in alternating order, which balances the propagation between the past and the future rounds.

A sequencer in every PE generates the slot address and the stage enables, and its counter triggers the growth step every \(K\) sweeps.
The schedule depends only on the number of sweeps since the start of decoding, so all sequencers run in lockstep without any synchronization signal.
Therefore, no control signal enters the PEs from outside the array during decoding.

\subsection{Commands, Completion, and Corrections}

\label{subsec:chain}

The host communicates with the array through the chains in Fig.~\ref{fig:array}(c), each of which advances one hop per clock cycle between adjacent PEs.
Commands enter at a corner to configure the PEs and load the defects, and every PE starts decoding after a fixed delay determined by its position, so all PEs begin the first sweep in the same cycle without any handshake.
Reset is also distributed along the command chain.
A PE is idle when no token, root parity, or activity flag remains in any of its slots or inbound links, and the array detects the completion by a flood fill in which every PE repeatedly combines its idle flag with those of its four neighbors.
The flood fill converges after \(2d\) cycles, every PE obtains the result at the same time, and one corner PE reports the completion to the host.
The correction bits then stream from PE to PE along the chains with a ready signal between neighbors only.

\section{Evaluation}

\subsection{Experimental Setup}

We evaluate Amatlas on the memory experiment of the rotated surface code generated by Stim~\cite{Gidney2021}.
Each circuit performs \(d-1\) rounds of stabilizer measurement followed by the readout of the data qubits, so the decoding graph has \(r=d\) syndrome rounds.
The noise model is that of CC~\cite{Barber2025}, which applies two-qubit depolarizing noise with probability \(p\) after every CNOT gate, single-qubit depolarizing noise with probability \(p/10\) to every other qubit in every time step, and measurement errors with probability \(p\).
Both types of stabilizer are decoded by one array, and the growth period is \(K=d+1\) unless stated otherwise.

We run four simulations on a cycle-accurate model verified against the RTL.
The first simulation obtains the logical error rate against \(p\) for \(d=5\) to \(21\), where each point stops at \(10^4\) logical errors or \(10^7\) shots.
The second obtains the logical error rate against the growth period \(K\) for \(d=9\), \(13\), and \(17\) at \(p=0.5\%\) with \(10^5\) shots per point.
The third obtains the distribution of cycle counts for \(d=3\) to \(25\) at \(p=0.1\%\) with \(10^6\) shots per distance.
The fourth obtains the distribution of cycle counts at \(p=0.5\%\), against \(K\) for \(d=9\), \(13\), and \(17\), and at \(K=d+1\) for \(d=21\) and \(25\), with \(10^5\) shots per point.
Error bars show 95\% Wilson intervals.

The RTL is implemented with Vivado 2024.2 for the Xilinx VCU129 FPGA (VU29P) as an out-of-context module, with a single clock constraint and the Performance\_Explore strategy.
We report the maximum clock frequency after place and route, and convert cycle counts to decoding times at that frequency.
Amatlas decodes the \(r\) rounds of a shot in one batch, and we define the decoding time per round as the decoding time of the shot divided by \(r\).

\subsection{Decoding Accuracy}

\label{subsec:accuracy}

\begin{figure*}[tbp]
  \centering
  \centerline{\includegraphics[width=\linewidth]{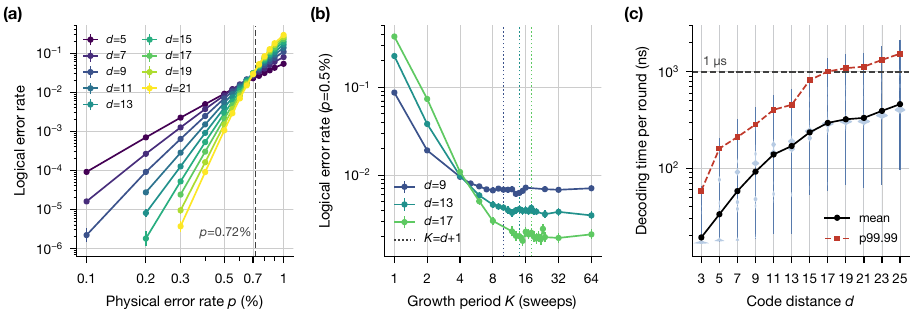}}
  \caption{Evaluation results under circuit-level noise. (a) Logical error rate for \(d=5\) to \(21\). The dashed line indicates the threshold of \(0.72\%\). (b) Reduction of the logical error rate with the growth period \(K\) at \(p=0.5\%\). The reduction saturates around \(K=d+1\). (c) Decoding time per round at \(p=0.1\%\) with the clock frequencies in Table~\ref{tab:fpga}. The width of each violin is proportional to the fraction of shots, and the vertical line spans the full range.}
  \label{fig:results}
\end{figure*}

Fig.~\ref{fig:results}(a) shows the logical error rate against \(p\) for \(d=5\) to \(21\).
The curves for \(d\ge 13\) cross between \(p=0.71\%\) and \(0.73\%\), and we estimate the threshold to be \(0.72\%\).
This value is close to the threshold of \(0.78\%\) reported for UF decoding under the same noise model~\cite{Barber2025}.

Fig.~\ref{fig:results}(b) shows the logical error rate against the growth period \(K\) for \(d=9\), \(13\), and \(17\) at \(p=0.5\%\).
Cluster information propagates at one hop per sweep and a cluster spans at most about \(d\) nodes, so merging converges within about \(d\) sweeps and the logical error rate is expected to saturate around \(K=d\).
The measured curves confirm this expectation at \(p=0.5\%\) as well as \(p=0.3\%\) and \(0.7\%\) (not shown), so the saturation point does not depend on \(p\).
We therefore set \(K=d+1\), where the logical error rate is indistinguishable from the saturated value within the statistical error.

\subsection{FPGA Implementation}

\label{subsec:impl}

\begin{table}[tbp]
\caption{Implementation Results and Decoding Time on the VCU129 FPGA}
\begin{center}
\setlength{\tabcolsep}{4pt}
\begin{tabular}{rrrrrrrr}
\toprule
\multirow{2}{*}{\(d\)} & \multirow{2}{*}{LUTs} & \multirow{2}{*}{\begin{tabular}[c]{@{}c@{}}LUT\\util.\end{tabular}} & \multirow{2}{*}{LUTRAMs} & \multirow{2}{*}{FFs} & \multirow{2}{*}{\begin{tabular}[c]{@{}c@{}}\(f_{\max}\)\\(MHz)\end{tabular}} & \multicolumn{2}{c}{Time/round (ns)} \\
\cline{7-8}
 & & & & & & mean & p99.99 \\
\midrule
3 & 14,304 & 0.8\% & 1,472 & 4,540 & 405 & 19 & 58 \\
5 & 46,552 & 2.7\% & 8,040 & 10,402 & 345 & 33 & 160 \\
7 & 88,966 & 5.2\% & 14,368 & 20,102 & 325 & 58 & 211 \\
9 & 155,620 & 9.0\% & 23,304 & 36,744 & 296 & 92 & 284 \\
11 & 242,560 & 14.0\% & 33,600 & 56,454 & 258 & 139 & 404 \\
13 & 340,709 & 19.7\% & 45,768 & 80,756 & 259 & 170 & 455 \\
15 & 462,357 & 26.8\% & 59,808 & 110,604 & 219 & 235 & 819 \\
17 & 666,567 & 38.6\% & 81,552 & 157,057 & 198 & 295 & 1,008 \\
19 & 849,293 & 49.1\% & 100,704 & 200,813 & 201 & 322 & 1,091 \\
21 & 1,074,093 & 62.2\% & 121,872 & 252,415 & 213 & 332 & 1,132 \\
23 & 1,294,021 & 74.9\% & 145,056 & 311,666 & 196 & 393 & 1,329 \\
25 & 1,575,375 & 91.2\% & 170,256 & 379,945 & 182 & 462 & 1,543 \\
\bottomrule
\end{tabular}
\label{tab:fpga}
\end{center}
\end{table}

Table~\ref{tab:fpga} lists the resource utilization, the clock frequency, and the decoding time per round at \(p=0.1\%\) for \(d=3\) to \(25\).
The array of \((d+1)^2\) PEs fits the FPGA up to \(d=25\), where it occupies 91\% of the LUTs.
The number of LUTs grows faster than the number of PEs, since the size of a PE increases with the number of slots and the width of the cluster identifiers.
LUTRAMs account for 10\% to 17\% of the LUTs, the design uses neither block RAMs nor DSP slices, and the flip-flop utilization stays below 11\%.

The clock frequency decreases from 405\,MHz at \(d=3\) to 182\,MHz at \(d=25\).
At every distance the critical path is confined to one PE or to two adjacent PEs, and in most cases it runs from the state memory of a PE through a link into the reduction register of the adjacent PE.
On this path the logic delay stays between 0.8 and 1.2\,ns, since the number of logic levels increases only from 7 to 11 with the width of the cluster identifiers, whereas the routing delay increases from 1.7 to 4.4\,ns.
The array fits in one super logic region (SLR) up to \(d=9\) and spans all four SLRs from \(d=17\).
At several distances from \(d=15\) the critical path is a link between adjacent PEs placed in different SLRs, and at \(d=25\) it is a link with a long route caused by the congestion at 91\% utilization.

\subsection{Decoding Time}

\begin{table}[t]
\caption{Average Decoding Time per Round \textnormal{(ns)} at Two Error Rates}
\begin{center}
\begin{tabular}{rrrrrr}
\toprule
\(d\) & 9 & 13 & 17 & 21 & 25 \\
\midrule
\(p=0.1\%\) & 92 & 170 & 295 & 332 & 462 \\
\(p=0.5\%\) & 148 & 269 & 516 & 649 & 953 \\
\bottomrule
\end{tabular}
\label{tab:ptime}
\end{center}
\end{table}

Fig.~\ref{fig:results}(c) shows the distribution of the decoding time per round at \(p=0.1\%\) for \(d=3\) to \(25\), and Table~\ref{tab:fpga} lists the mean and the 99.99th percentile.
The average number of cycles per round grows almost linearly with \(d\).
The number of growth steps in a decoding at \(p=0.1\%\) does not depend on \(d\), and each growth period consists of \(K=d+1\) sweeps of \(2r+1\) cycles, so the number of cycles per decoding grows with \(rd\) and the number per round with \(d\).
The number of cycles also increases linearly with \(K\), as measured for \(d=9\), \(13\), and \(17\) at \(p=0.5\%\).
Together with the decrease of the clock frequency described in Section~\ref{subsec:impl}, the average decoding time per round increases from 19\,ns at \(d=3\) to 462\,ns at \(d=25\) and stays below 1\,\textmu s at every distance.
The 99.99th percentile grows in proportion to the mean and exceeds 1\,\textmu s for \(d\ge 17\).
As Table~\ref{tab:ptime} shows, raising \(p\) from \(0.1\%\) to \(0.5\%\) increases the average decoding time by a factor of 1.6 to 2.1, and it remains below 1\,\textmu s up to \(d=25\).

\subsection{Comparison With Prior Work}

\label{subsec:comparison}

\begin{table}[tbp]
\caption{Comparison With FPGA Union-Find Decoders Under Circuit-Level Noise}
\begin{center}
\setlength{\tabcolsep}{4pt}
\begin{tabular}{lccccr}
\toprule
Decoder & PEs & Stage sync. & FPGA & Max \(d\) & Time/round \\
\midrule
CC~\cite{Barber2025} & 1 & No & VU3P & 23 & 1.18\,\textmu s \\
Helios~\cite{Liyanage2024} & \(O(d^3)\) & Yes & VU29P & 17 & 23.7\,ns \\
Snowflake~\cite{Tahaab2026} & \(O(d^3)\) & Yes & KU5P & 9 & 1.06\,\textmu s \\
LCD~\cite{Ziad2025} & \(O(d^2)\) & Yes & VU19P & 17 & 460\,ns \\
Amatlas (ours) & \(O(d^2)\) & No & VU29P & 25 & 462\,ns \\
\bottomrule
\end{tabular}
\label{tab:comparison}
\end{center}
\end{table}

Table~\ref{tab:comparison} compares Amatlas with the FPGA implementations of UF decoding under circuit-level noise.
The table lists the largest reported distance on each FPGA and the average decoding time per round at that distance and at \(p=0.1\%\).
The comparison is not on equal terms.
CC, Helios, Snowflake, and LCD decode the graph of only one type of stabilizer, whereas Amatlas decodes both types in one array, so the resources of Amatlas cover twice the number of nodes.
Snowflake implements the unrotated surface code under a different circuit-level noise model, and LCD is evaluated under a noise model that adds leakage.
Helios-\(n\)~\cite{Liyanage2024} reaches \(d=51\), but its implementation has been reported only under phenomenological noise, and it is therefore not listed.

In Amatlas the logic of a PE is shared among the \(r\) slots, so the number of LUTs grows more slowly than the number of nodes.
Helios reaches \(d=17\) at 75\,MHz with 889K LUTs, whereas Amatlas reaches \(d=25\) at 182\,MHz on the same FPGA and occupies 667K LUTs at \(d=17\) while decoding both types of stabilizer.
At \(d=9\), Amatlas occupies 156K LUTs, whereas the full configuration of Snowflake requires 238K LUTs for one type of stabilizer and does not fit its FPGA, so the implemented configuration is optimized for area and runs at 60\,MHz.
LCD occupies 252K LUTs at \(d=17\), less than half of Amatlas, and CC occupies only 21.7K LUTs at \(d=23\).

In decoding time, only Amatlas stays below 1\,\textmu s per round on average beyond \(d=21\) at \(p=0.1\%\).
At \(d=17\), Amatlas decodes in 295\,ns per round against 460\,ns for LCD.
At \(p=0.5\%\), CC stays below 1\,\textmu s only up to \(d=11\)~\cite{Barber2025}, whereas Amatlas stays below 1\,\textmu s up to \(d=25\).

\section{Discussion}

\label{sec:discussion}

Under circuit-level noise, a fault on an ancilla qubit propagates to two data qubits and produces the hook edge in Fig.~\ref{fig:surface_code}(c), which connects two stabilizers of the same type in adjacent rounds across a cell.
The links of Amatlas connect only diagonally adjacent PEs, so we decompose every hook edge into a path of two edges when constructing the decoding graph.
The threshold in Section~\ref{subsec:accuracy} indicates that the decomposition costs little accuracy, and the remaining gap to MWPM decoding comes from the uniform growth of UF decoding itself.
Weighted growth may narrow this gap, and we expect it to fit the token-based formulation without any change to the links.

The clock frequency is limited by the routing between adjacent PEs and not by the logic, so it can be raised without changing the architecture.
The critical path in Section~\ref{subsec:impl} contains both the link and the first stage of the reduction, and the RTL already provides a synthesis parameter that moves the reduction into the select stage, which would balance the logic between the two cycles at large distances.
A floorplan that aligns the rows of the array with the SLR boundaries would remove the crossings observed from \(d=15\).
On an ASIC, the regular array maps directly onto the layout and the routes between adjacent PEs are short and uniform, so we expect the frequency to depend only weakly on \(d\), although the number of cycles per round still grows with \(d\).

\section{Conclusion}

We presented Amatlas, a two-dimensional array architecture for UF decoding without global coordination.
Each PE stores the decoding state of all measurement rounds in its local memory, and a residual-syndrome invariant allows growth to proceed on a fixed schedule without stage synchronization.
The RTL implementation fits the VCU129 FPGA up to \(d=25\) with an average decoding time below 1\,\textmu s per round under circuit-level noise, and the threshold of 0.72\% is close to that of prior UF decoders.

\section*{Acknowledgment}
This work was supported in part by JST Moonshot R\&D Grant Number JPMJMS256F and JST SPRING, Grant Number JPMJSP2108.
We used Claude (Anthropic) for manuscript revision and coding assistance.

\bibliographystyle{IEEEtran}
\bibliography{references}

\end{document}